# Gamma ray Large Area Space Telescope (GLAST) Balloon Flight Engineering Model: Overview

D. J. Thompson, G. Godfrey, S. Williams, J. E. Grove, T. Mizuno, H. F.-W. Sadrozinski, *Senior Member, IEEE*, T. Kamae, J. Ampe, Stuart Briber, James Dann, E. do Couto e Silva, R. Dubois, Y. Fukazawa, B. Giebels, G. Haller, T. Handa, R. C. Hartman, K. Hirano, M. Hirayama, R. P. Johnson, W. N. Johnson, A. Kavelaars, H. Kelly, Steve Kliewer, T. Kotani, J. Krizmanic, W. Kroger, M. Kuss, D. Lauben, T. Linder, M. Lovellette, N. Lumb, Joe Manildi, P. Michelson, H. Mizushima, A. Moiseev, P.L. Nolan, S. Ogata, J. F. Ormes, M. Ozaki, G. Paliaga, B. Phlips, S. Ritz, L. S. Rochester, F. M. Roterman, W. A. Rowe, J. J. Russell, R. Schaefer, T. Schalk, D. Sheppard, S. Singh, M. Sjogren, G. Spandre, T. Usher, P. Valtersson, A. P. Waite, J. Wallace, A. Webster, and D. Wood, *on behalf of the GLAST Large Area Telescope Collaboration*

***Abstract* -- The Gamma Ray Large Area Space Telescope (GLAST) Large Area Telescope (LAT) is a pair-production high-energy (>20 MeV) gamma-ray telescope being built by an international partnership of astrophysicists and particle physicists for a satellite launch in 2006, designed to study a wide variety of high-energy astrophysical phenomena. As part of the development effort, the collaboration has built a Balloon Flight Engineering Model (BFEM) for flight on a high-altitude scientific balloon. The BFEM is approximately the size of one of the 16 GLAST-LAT towers and contains all the components of the full instrument: plastic scintillator anticoincidence system (ACD), high-Z foil/Si strip pair-conversion tracker (TKR), CsI hodoscopic calorimeter (CAL), triggering and data acquisition electronics (DAQ), commanding system, power distribution, telemetry, real-time data display, and ground data processing system. The principal goal of the balloon flight was to demonstrate the performance of this instrument configuration under conditions similar to those expected in orbit. Results from a balloon flight from Palestine, Texas, on August 4, 2001, show that the BFEM successfully obtained gamma-ray data in this high-background environment.**

Manuscript received November 6, 2001; revised March 15, 2002.

This work is supported in part by Department of Energy contract DE-AC03-76SF00515 and NASA contract NAS598039.

D. J. Thompson, R. C. Hartman, H. Kelly, T. Kotani, J. Krizmanic, A. Moiseev, J. F. Ormes, S. Ritz, R. Schaefer, D. Sheppard, S. Singh are with NASA Goddard Space Flight Center, Greenbelt, MD 20771 USA (telephone: 301-286-8168, e-mail: djt@egret.gsfc.nasa.gov).

G. Godfrey, E. do Couto e Silva, R. Dubois, B. Giebels, G. Haller, T. Handa, T. Kamae, A. Kavelaars, T. Linder, L. S. Rochester, F. M. Roterman, J. J. Russell, T. Usher, A. P. Waite are with the Stanford Linear Accelerator Center

S. M. Williams, D. Lauben, P. Michelson, P.L. Nolan, J. Wallace are with Stanford University

T. Mizuno, Y. Fukazawa, K. Hirano, H. Mizushima, S. Ogata are with Hiroshima University

J. E. Grove, W. N. Johnson, M. Lovellette, B. Phlips, D. Wood are with the Naval Research Laboratory

H. f.-W. Sadrozinski, M. Hirayama, R. P. Johnson, W. Kroger, G. Paliaga, W. A. Rowe, T. Schalk, A. Webster are with the University of California, Santa Cruz

M. Kuss, N. Lumb, G. Spandre are with INFN-Pisa and University of Pisa

M. Ozaki is with the Stanford Linear Accelerator Center and the Institute of Space and Astronautical Science

M. Sjogren and P. Valtersson are with the Stanford Linear Accelerator Center and the Royal Institute of Technology

J. Ampe is with the Naval Research Laboratory and Praxis

Stuart Briber is with Independence HS, San Jose CA

James Dann is with St. Ignatius HS, San Francisco CA

Steve Kliewer is with Paso Robles HS, Paso Robles CA

Joe Manildi is with Watsonville HS, Watsonville CA

## I. Introduction

The Gamma Ray Large Area Space Telescope (GLAST), planned for launch by NASA in 2006, carries two successors to instruments on the Compton Gamma Ray Observatory. The GLAST Burst Monitor extends the work of the Burst and Transient Source Experiment, while the Large Area Telescope (LAT) represents a significant advance over the Energetic Gamma Ray Experiment Telescope (EGRET).

The GLAST LAT is a pair-production high-energy (E> 20 MeV) gamma-ray telescope [1]. Its scientific objectives include revealing high-energy processes of active galactic nuclei and their jets, extragalactic and galactic diffuse emissions, dark matter, supernova remnants, pulsars, and the unidentified high energy gamma-ray sources.

As part of the LAT development effort, the collaboration has built and flown on a balloon a functional prototype of one of the 16 LAT towers, called the Balloon Flight Engineering Model (BFEM). This paper presents an overview of the balloon test program.

## II. Rationale and Goals for the BFEM

Although the GLAST LAT has been developed using extensive simulations and beam tests [2]-[3], it was recognized that a balloon flight could provide a system-level test under near spaceflight conditions. In particular, operating successfully in the atmospheric background with its mix of particles and photons arriving from all directions randomly in time at a high rate is a test that adds further confidence that the design approach of the LAT will work successfully in space. Such a test was mandated by the GLAST Announcement of Opportunity from NASA.

Four specific objectives were adopted for the balloon flight:

Goal 1) Validate the basic LAT design at the single tower level under flight conditions.

Goal 2) Show the ability to take data in the high isotropic background flux of energetic particles in the balloon environment.

Goal 3) Record all or partial particle incidences in an unbiased way that can be used as a background event data base.

Goal 4) Find an efficient data analysis chain that meets the requirement for the future Instrument Operation Center of the GLAST LAT.

## III. Planning and Design Approach

Engineering design for a balloon flight falls between that of ground testing and that required for a satellite. Commercial electronics can often be used, but they must operate in a remote and space-like environment (constrained power source, near-vacuum, cold surroundings, but full sun exposure for a daytime flight). In order to make the best use of the balloon flight data and have the minimum distraction from the satellite development, the balloon program had to be carried out quickly and with minimum resources. This goal was achieved for the BFEM by using a large amount of existing hardware: the detectors were those used by the GLAST LAT collaboration for an accelerator beam test (the Beam Test Engineering Model) [3] with some modifications. Much of the supporting hardware was borrowed from other balloon flight programs - a pressure vessel (needed because the prototype electronics were not designed to operate in a vacuum), a gondola to hold the instrument and connect to the balloon and parachute (with safety margin), and some of the interface electronics to handle commands and data transfer [4] between the instrument and the National Scientific Balloon Facility (NSBF) telemetry system. As for the GLAST/LAT project itself, the BFEM development was a collaborative effort. Table I shows how the various institutions contributed to specific parts of the program.

Table I
Division of Responsibility for BFEM Development

| Organization | Responsibility |
|---|---|
| NASA/National Scientific Balloon Facility | Balloon, parachute, rigging, batteries, command/data electronics, launch support, recovery support |
| NASA/Goddard Space Flight Center | Gondola, pressure vessel, anticoincidence detector, magnetometer, interface electronics, assembly, test, data analysis |
| Stanford Linear Accelerator Center | Pressure vessel modification, cooling system, on-board software, assembly, test, data handling/analysis |
| Stanford University | Data Acquisition System, housekeeping, electrical ground support equipment, assembly, test, data analysis |
| Hiroshima University | External gamma targets, simulations, data analysis |
| Naval Research Laboratory | Calorimeter, Balloon Interface Unit, command and on-board software, assembly, test, data analysis |
| University of California, Santa Cruz | Tracker, recovery support, data analysis |
| INFN-Pisa and University of Pisa | Event display |

## IV. Integration and Test

Many details of the beam test instrument on which the BFEM is based, including calibrated parameters such as point spread function and energy resolution, are described in [3]. The basic elements of the detector (shown schematically in Fig. 1) are:

- A 13-segment plastic scintillator anticoincidence detector (ACD) with waveshifting fiber and photomultiplier readout, designed to help separate the enormous charged particle background from the gamma rays.
- A Si-strip tracker (TKR) with 13 x-y layers (32 cm x 32 cm area, although the top five layers were not fully instrumented with Si strips), interleaved with thin lead foils (eight 3.5% radiation length foils at the top, three 25% radiation length foils, plus two layers with no foils at the bottom) to provide pair production converters. The tracker, which provides the instrument trigger and measures trajectories of particles, is read out by custom electronics.
- A segmented (80 logs, each 3.0 cm x 2.3 cm x 31 cm) hodoscopic (eight layers of 10 logs in alternating x and y directions) CsI(Tl) calorimeter (CAL) for energy measurement, read out by photodiodes.
- A set of external gamma target (XGT) plastic scintillators read out by phototubes, located above the rest of the instrument and designed to provide notification of cosmic ray interactions of potential interest.
- A software-based data acquisition system (DAQ) to configure the detectors, assemble data from the subsystems, and then record/send the data.

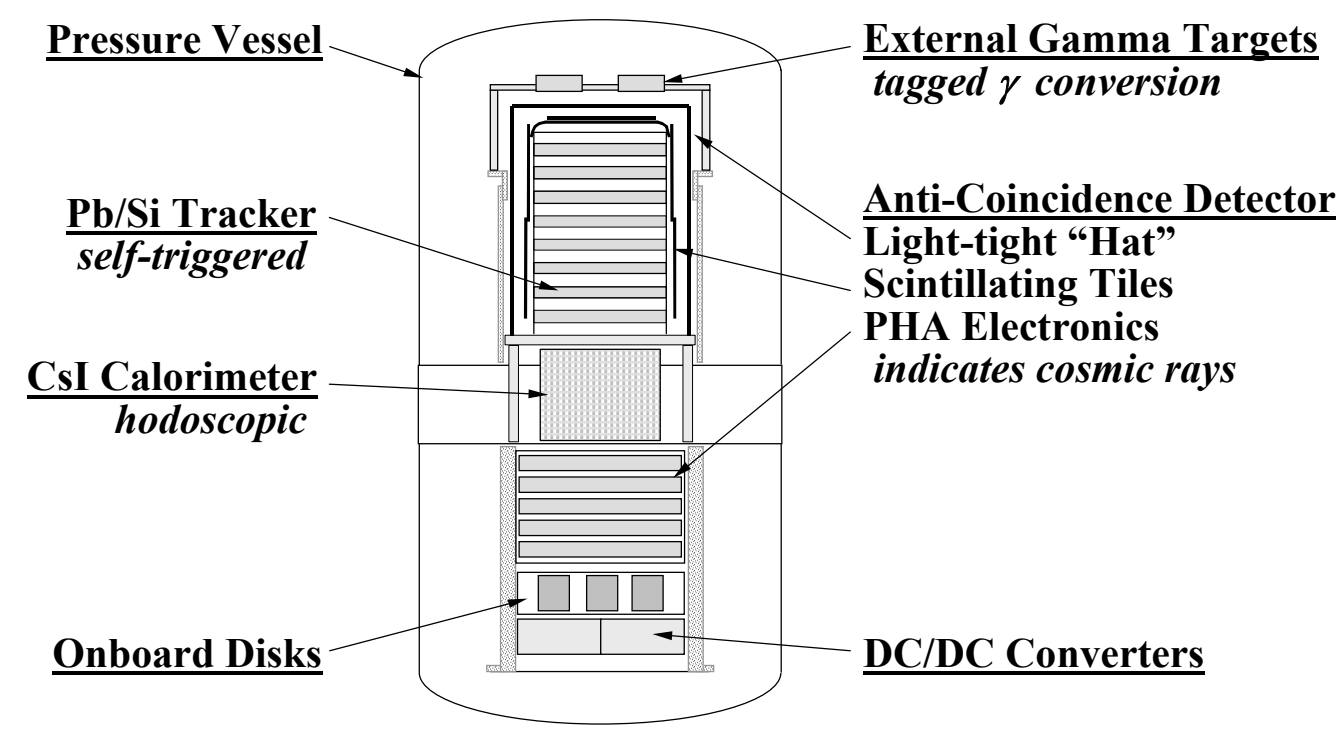

Fig. 1 - Schematic diagram of the Balloon Flight Engineering Model. The detector elements are described in the text, with additional details in [3].

Several other essential elements of the BFEM were:

- A Balloon Interface Unit (BIU) to handle the interfaces for commands and telemetry between the DAQ and the NSBF instrument package.
- Electrical ground support equipment (EGSE) to send commands and display real-time telemetry.
- Temperature, voltage, current, pressure, and magnetic field sensors to provide housekeeping information, along with a Global Positioning System device.

Fig. 2 shows schematically the electronics components and data flow for the BFEM. In order to record an unbiased sample of data, the trigger consisted of signals in any three consecutive x-y tracker layers (six-fold coincidence). At the time of a trigger, signals from the entire TKR, ACD, CAL, and XGT assembly were recorded.

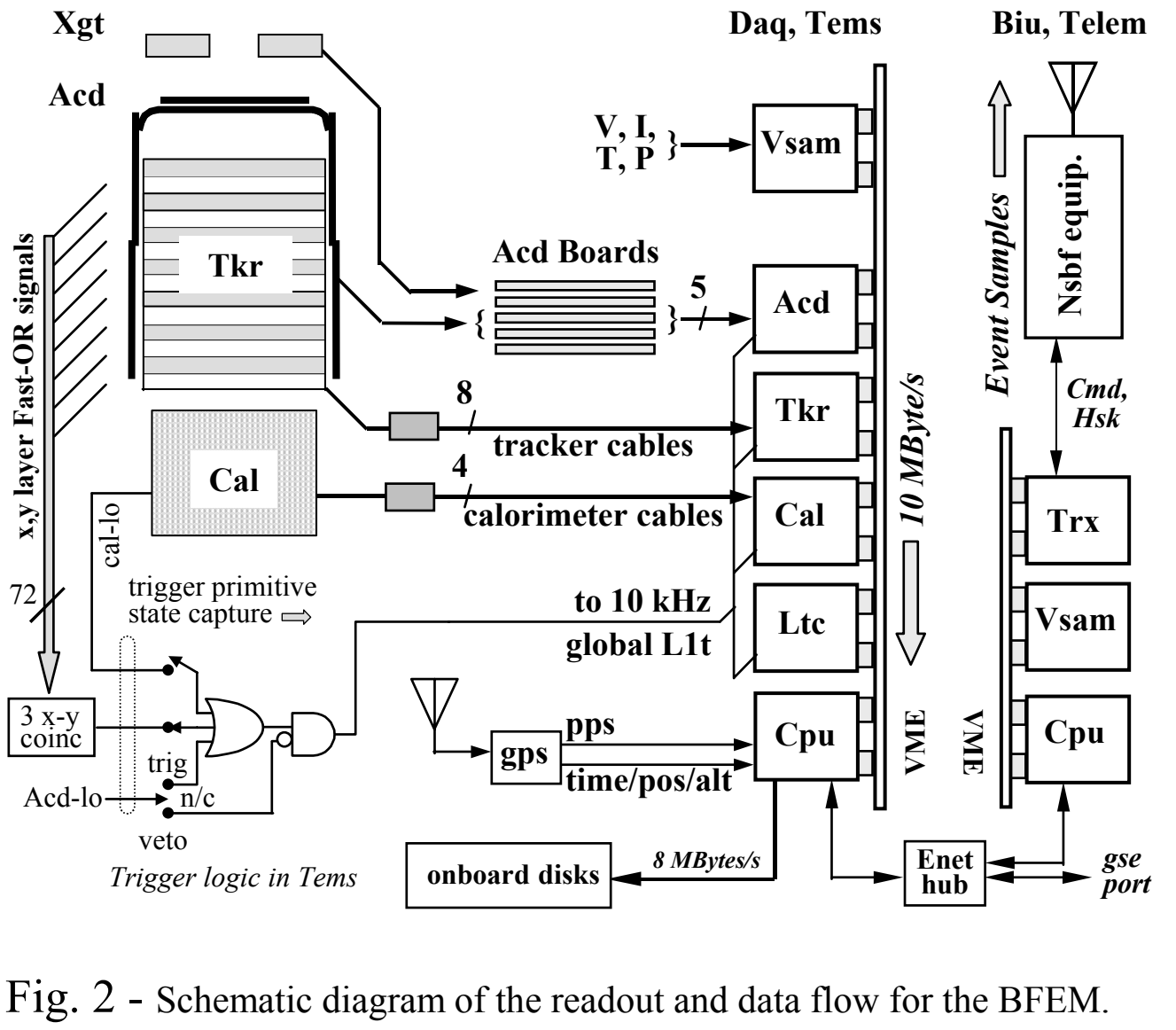

Fig. 2 - Schematic diagram of the readout and data flow for the BFEM.

The basic elements of the BFEM were assembled and tested at the Stanford Linear Accelerator Center (SLAC) in January-May, 2001. An example of the performance of the BFEM is shown in Fig. 3, taken from the EGSE display. This cosmic ray track penetrates the BFEM and is seen by all the detectors.

The BFEM was shipped to Goddard Space Flight Center (GSFC) in May, 2001. There the remaining housekeeping detectors were added, the BIU was completed, the onboard software was upgraded to handle autonomous operation, several additional real-time displays were added, and the instrument was mounted into its flight gondola. Extensive testing at both SLAC and GSFC suggested a possible thermal problem, and so several modifications were made to allow better cooling, including fans and a radiator that was fed from the outside by chilled water during ground operations. Following a review by a scientific team with considerable balloon experience, the BFEM was shipped in July, 2001, to the National Scientific Balloon Facility in Palestine, Texas.

## V. Operations at NSBF

With substantial critical support from the staff of NSBF, we completed preparations for the balloon flight: batteries were wired and tested, electrical and mechanical interfaces with the NSBF equipment were checked, insulation and crush pads were added, the pressure vessel was leak-tested, and further instrument tests and calibrations were carried out. Fig. 4 shows the fully-assembled BFEM hanging from the NSBF "Tiny Tim" launch vehicle during testing. Following a flight readiness review on Aug. 3, the BFEM was launched on Aug. 4 using a 29 million cubic ft. (800,000 cubic m) balloon. After a 2 hour ascent to an altitude of 38 km (atmospheric depth 3.8 g/cm$^2$), the balloon was carried rapidly west. It reached the limit of telemetry after three hours at float altitude, and the flight was terminated. The BFEM was recovered (after a fairly rough descent and landing) near San Angelo, Texas. The total time from the start of the GLAST BFEM development to launch was about 13 months, thus achieving the goal of a rapid completion.

## VI. Results

Even before the flight had been completed, the BFEM demonstrated that the first three goals of the mission had been achieved:

1) The detectors worked well throughout the flight. The trigger, based on three x-y signals from consecutive layers of the tracker, operated successfully. The tracker-based trigger was an important departure from previous gamma-ray telescopes. The basic concept of the LAT was validated.
2) The high atmospheric background proved no obstacle to the BFEM data collection. Even through the Pfotzer maximum, the trigger rate never exceeded 1.5 KHz, well below the 6 KHz that the BFEM could handle. The rate at float altitude was 500 Hz. The trigger rate as a function of altitude is shown in Fig. 5.
3) A wide variety of event types was seen. Although the vast majority of triggers were cosmic rays as expected, some showers and gamma-ray pair production events were seen, along with a number of "short-track" events that require further analysis. The data certainly provide a reference set of triggers that are being used to compare and calibrate the simulations for both the BFEM and the flight unit.

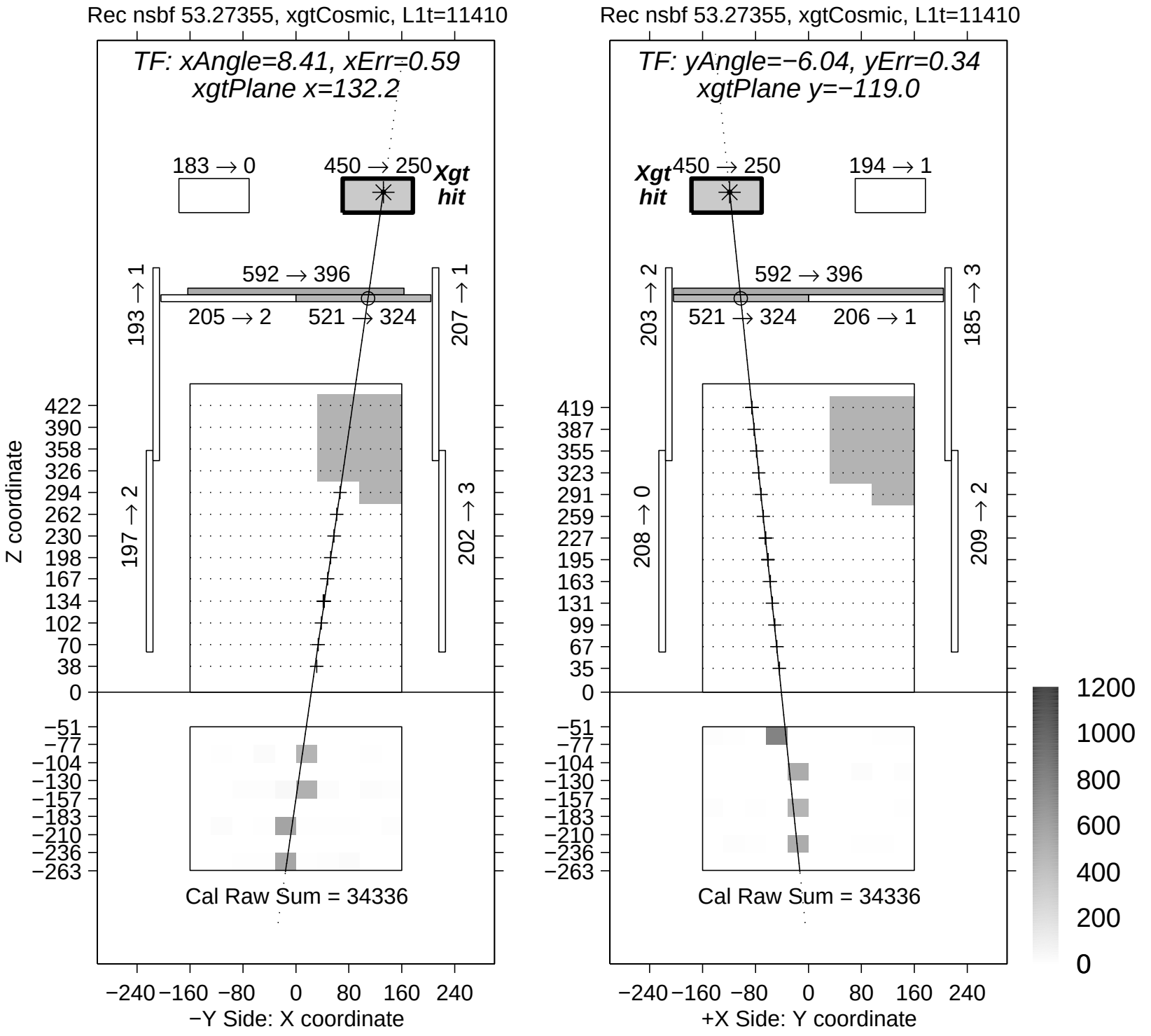

Fig. 3 - EGSE display (two views) of a cosmic ray event in the BFEM. The tracker layers are shown as dotted lines in the middle of the picture, with the non-instrumented area shaded. The tracker layers hit are shown by the + symbol. In the XGT, the ACD, and the CAL, the shading is a measure of the energy deposit. In the CAL, the ends of the logs with energy deposit are shown.

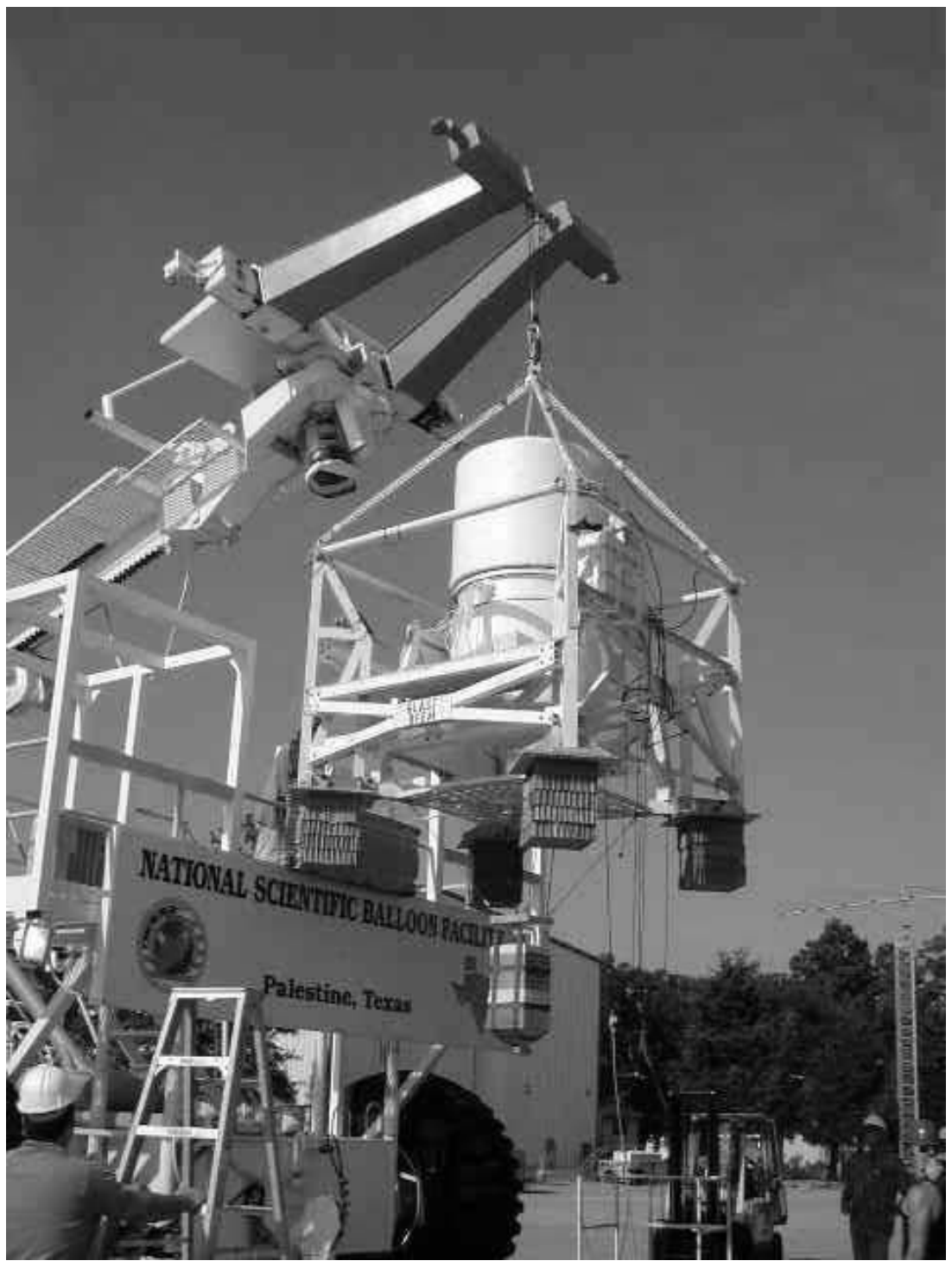

Fig. 4 - The GLAST BFEM during testing at NSBF, Palestine, Texas.

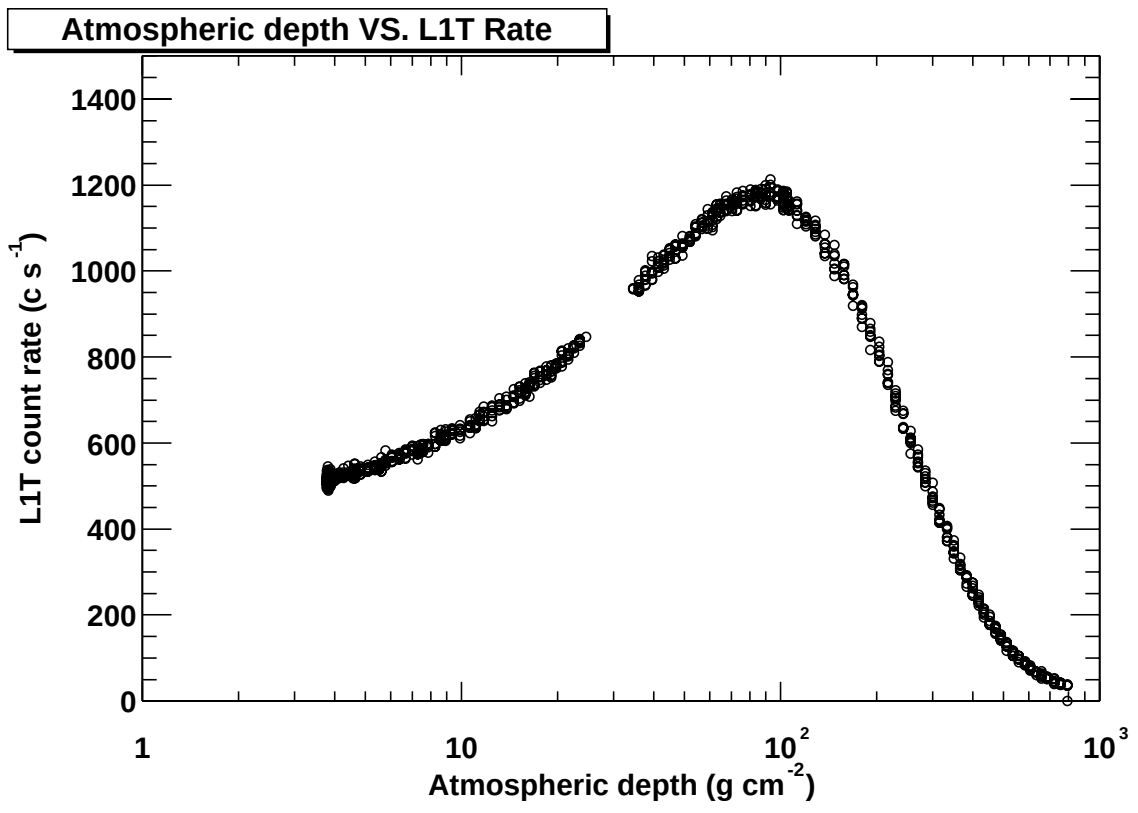

Fig. 5 - BFEM trigger rate as a function of atmospheric depth. The gap in data near atmospheric depth 30 g/cm$^2$ resulted from the reconfiguration of the system to shut down the on-board disks.

The detector subsystems all performed as expected. Although some tracker readout problems had been seen on the ground under high rate conditions and the on-board software had been modified to handle such conditions, these problems did not occur during flight. The tracker performance seen in Fig. 3, with essentially 100% efficiency and no significant noise, was characteristic of the events seen in flight. Using the tracker as a guide, the ACD was able to construct pulse height distributions for each of the tiles using tracks that were likely to penetrate the tile. Fig. 6 shows one of those, which shows the characteristic Landau distribution. The lower end of the distribution lies well above any noise from the phototube, giving confidence that the efficiency of the ACD is high. The calorimeter also used the tracker information to identify penetrating particles, and from those tracks was able to construct a pulse height distribution showing not only singly-charged particles but also some cosmic ray helium particles, as shown in Fig. 7.

The fourth goal of the BFEM mission, the development of a demonstration data analysis system, started well before the balloon launch. The work was carried out in parallel with continued development of the satellite data system. Some additional information about the data system is given in [5]. The processing of the data followed the planned pattern of the flight program, with conversion to a ROOT format, subsystem analyses to determine in-flight calibrations, and pattern recognition (RECON) to categorize the events. An event display for use with both the simulation data and the flight data was developed. Due to the limited quantity of data (the one disappointment during the balloon flight was a leak in the pressure vessel that forced the shut-down of onboard disks that would have collected a much larger volume of data), sophisticated cataloging and retrieval methods were not needed.

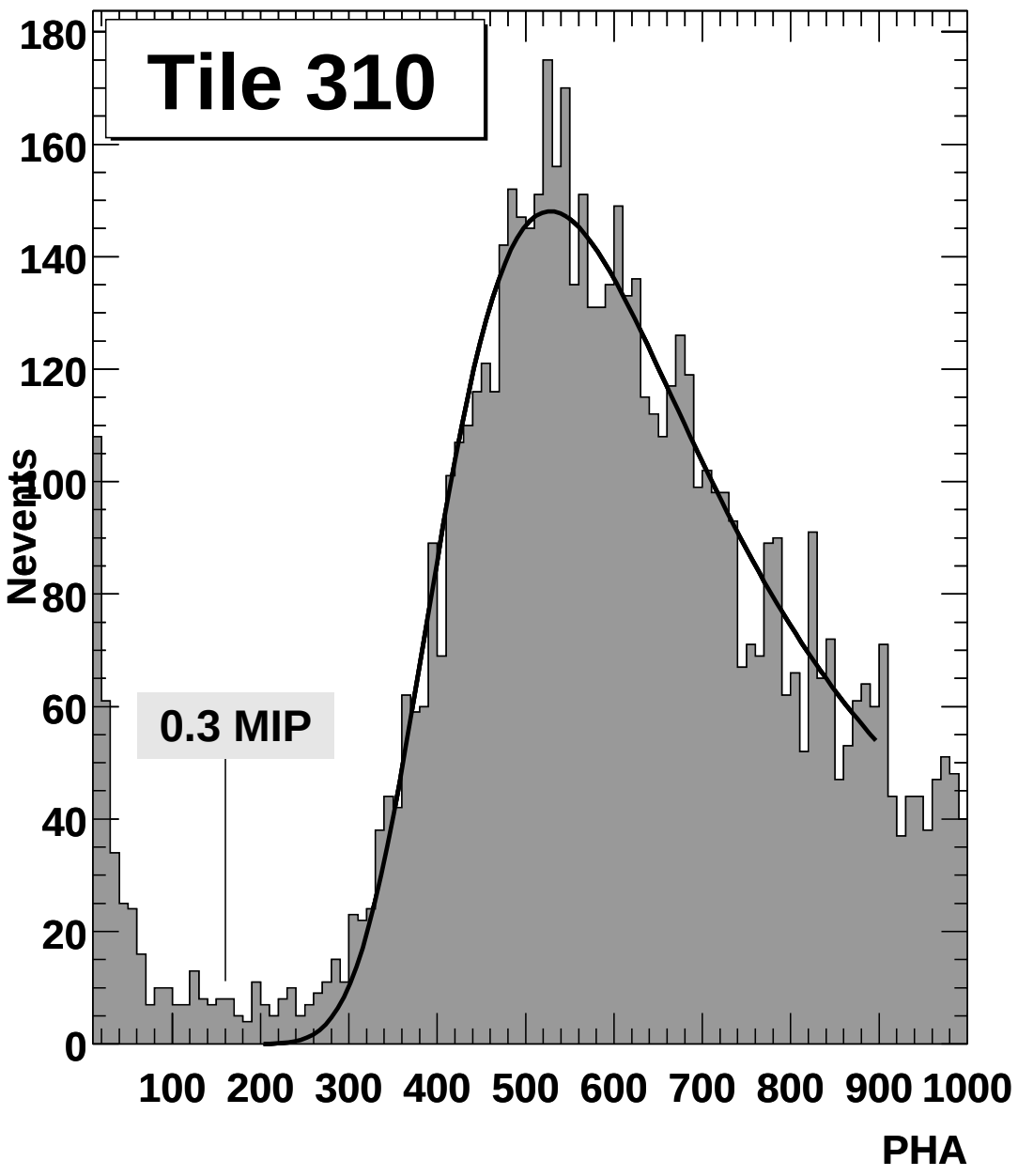

Fig. 6 - Pulse height distribution for charged-particle signals in one anticoincidence scintillator tile. At the nominal threshold of 0.3 MIP (Minimum Ionizing Particle), most of the triggers produce a signal.

Based on screening criteria developed using simulations of the LAT and past experience with EGRET, a set of event selections was developed for the balloon configuration. For the data taken during the float portion of the flight, these selections reduced the data from the 100,000 triggers recorded to fewer than 300 candidate events, consistent with the expectation from simulations that identifiable atmospheric gamma rays represent fewer than 1% of the triggers. A visual examination of these events shows that they are largely consistent with being gamma-ray pair production events as expected. A sample is shown in Fig. 8. Although data analysis will continue to refine the results, the basic conclusion is that a workable data system does exist, thus fulfilling the fourth goal of the balloon flight program.

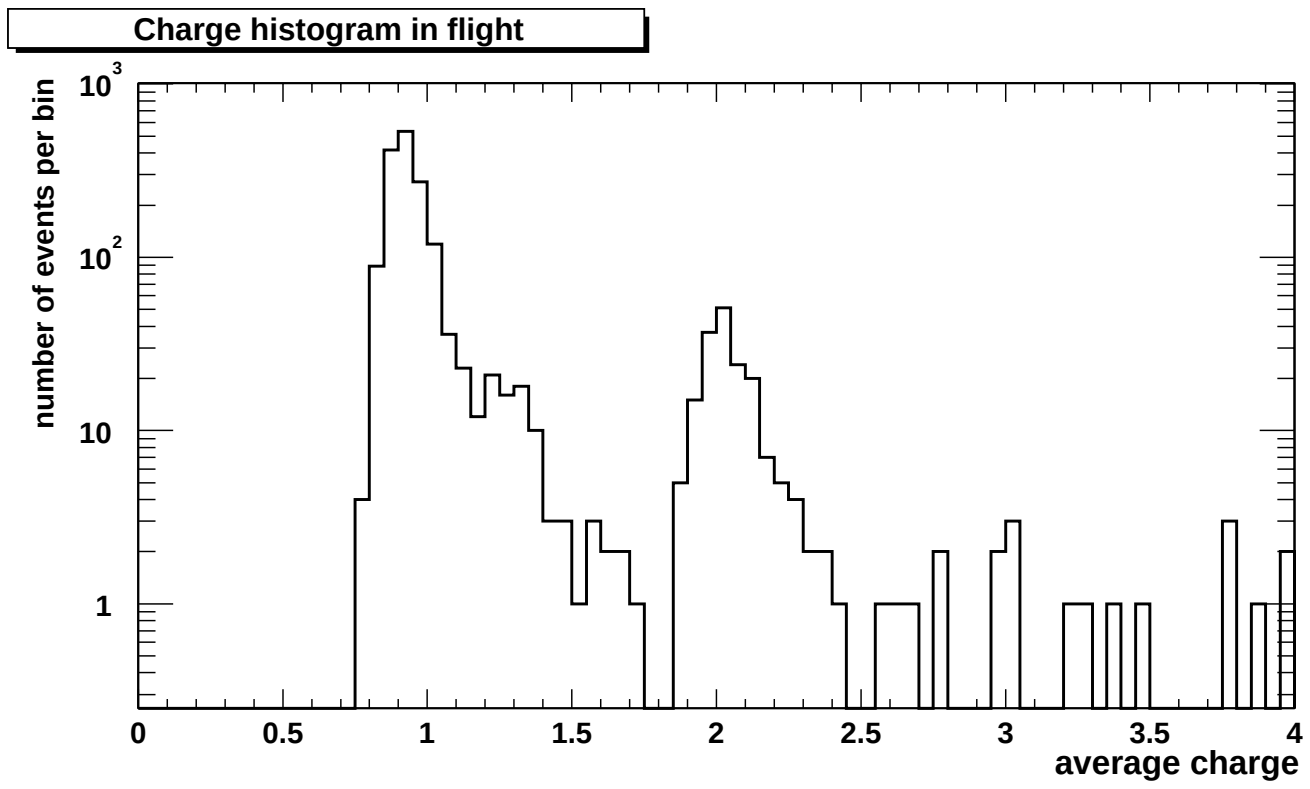

Fig. 7 - Charge histogram derived from pathlength-corrected total energy deposition in Calorimeter. Events were selected by requiring that the charge measurement in each layer be consistent with the average charge. The charge scale was established through electronic and sea-level muon calibration. Pathlength corrections were derived from Tracker trajectories.

A comparison of the observed trigger rate with that modeled for the BFEM using a GEANT4-based simulation [6] is shown in Table II. The reasonable agreement for both total triggers and "neutral" events (ones with no measurable energy deposit in the ACD) is further indication that the BFEM performed as expected.

TABLE II

COMPARISON OF MODELED AND OBSERVED TRIGGER RATES

| Trigger Type | Modeled (Hz) | Observed (Hz) |
|---|---|---|
| All Triggers | 540 | 500 |
| Neutral Triggers | 65 | 50 |

The balloon flight also provided two unanticipated tests. The leak in the pressure vessel forced the detectors to operate in a fairly low pressure, where convection could no longer be relied on for thermal control. All the subsystems continued to operate. The very rough descent and landing also stressed the instrument. Shocks exceeding 20g were recorded, but none of the detectors suffered noticeable damage. The tracker in particular was operated after the flight with no measurable change in performance.

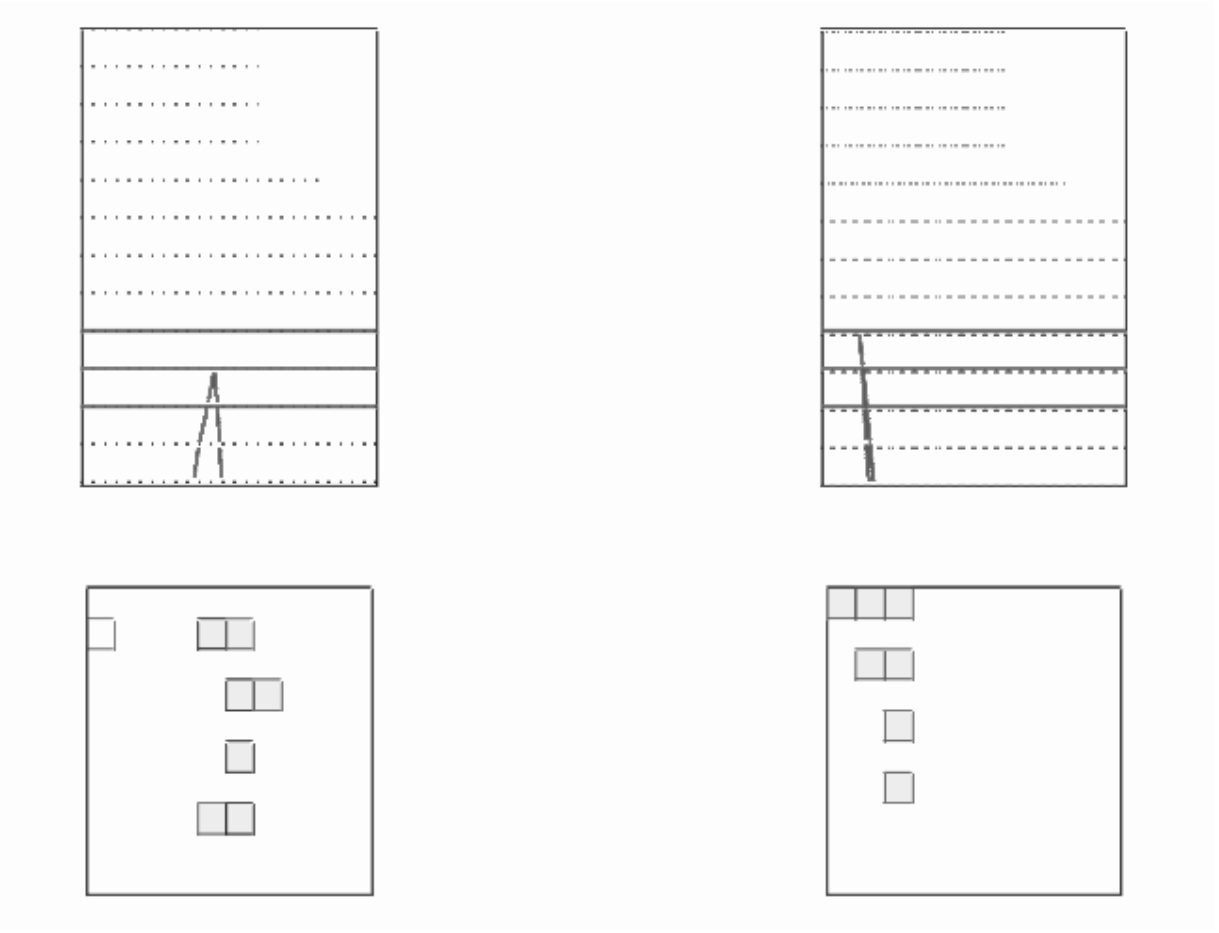

Fig. 8 - Orthogonal views of a pair production event in the BFEM. This low-resolution screen display is used to check the performance of the data processing. No ACD signals were present, as expected for a gamma-ray. The tracks from the pair production event can be seen in both the tracker (upper boxes) and the calorimeter (lower boxes). In the tracker, the dotted lines in the upper section are layers with thin converters (the blank region being the uninstrumented section), the solid horizontal lines are the layers with thick converters, and the lower dotted lines are the layers with no converters. In the calorimeter, the shaded boxes show the ends of the CsI logs in which energy was deposited. The lines showing the tracks in the tracker are those assigned by RECON, the pattern recognition program.

## VII. FUTURE WORK

The availability of the data collected during the flight and the simulations developed to model the BFEM offer opportunities to carry out analysis beyond this demonstration that the BFEM met its basic goals. Future work will include:

- Comparing details of the model (distribution of tracker layer hits, angular distribution of charged and neutral events, energy deposits, etc.) with what was seen during the flight, and determining what parameters (such as assumed primary and secondary cosmic ray spectra for various particle types) could be adjusted (within observational uncertainties) to produce better agreement [6].
- Improving event selection techniques using both the flight data and the simulations, and then using those criteria to derive improved response functions for the BFEM, including absolute effective area as a function of energy and angle [7].
- Using the optimized simulations to construct an atmospheric gamma-ray spectrum and angular distribution, which can be compared to previous balloon data.
- Comparing the techniques derived for the BFEM analysis with those being developed for the flight unit in order to highlight possible improvements.

## VIII. ACKNOWLEDGEMENT

We thank the staff of the National Scientific Balloon Facility, Palestine, Texas, for excellent support of the launch, flight, and recovery.

## IX. REFERENCES

[1] N. Gehrels and P. Michelson, "GLAST: the next-generation high energy gamma-ray astronomy mission," *Astroparticle Physics*, Vol. 11, Issue 1-2, pp. 277-282, 1999.

[2] W. B. Atwood, S. Ritz, P. Anthony, E. D. Bloom, P. F. Bosted, J. Bourotte, et al., "Beam test of gamma-ray large area space telescope components," *Nucl. Instr. Meth.* , Vol. A446, pp. 444-460, 2000.

[3] E. do Couto e Silva, G. Godfrey, P. Anthony, R. Arnold, H. Arrighi, E. Bloom, et al., "Results from the beam test of the engineering model of the GLAST Large Area Telescope," *Nucl. Instr. Meth*., Vol A474/1, pp 19-37, 2001.

[4] E. Christian, "A Telemetry/Frame Sync board for use in balloon payloads," *Proceedings of the 24th International Cosmic Ray Conference (Rome),* Vol. 3, pp. 623-626, 1995.

[5] T. H. Burnett, A. Chekhtman, E. do Couto e Silva , R. Dubois, D. Flath, I. Gable, et al., "Gamma ray large area space telescope (GLAST) balloon flight data handling overview," *IEEE Trans. Nucl. Sci*., Vol 49, Aug. 2002.

[6] T. Mizuno, T. Kamae, T. Handa, T. Lindner, H. Mizushima, S. Ogata,, et al., " Study of data from the GLAST balloon prototype based on a cosmic ray and instrument simulator," in preparation.

[7] T. Kotani, R. C. Hartman, and J. F. Ormes, "Event screening and selection on data from the GLAST balloon prototype," in preparation.